\documentclass[a4paper,12pt]{article}

\usepackage[utf8]{inputenc}
\usepackage{makeidx}         
\usepackage{graphicx}        
\usepackage{xcolor}
\usepackage{amsmath}

\def\pertp{\varepsilon}

\def\gfamp{g}

\def \Sf {\Sigma_0^+}
\def \Sv {\Sigma_0^-}
\def \Sb {\Sigma_0}

\def \tf {t_+}
\def \rf {r_+}
\def \thetaf {\theta_+}
\def \phif {\varphi_+}
\def \Rf {R_+}

\def \nuf {\nu^+}
\def \lf {\lambda^+}
\def \of {\tilde{\omega}^+}
\def \mf {m_0^+}
\def \hf {h_0^+}
\def \jf {j^+}
\def \Mf {M^+}

\def \tv {t_-}
\def \rv {r_-}
\def \thetav {\theta_-}
\def \phiv {\varphi_-}
\def \Rv {R_-}
\def \jv {j^-}
\def \Mv {M^-}

\def \nuv {\nu^-}
\def \lv {\lambda^-}
\def \ov {\tilde{\omega}^-}
\def \mv {m_0^-}
\def \hv {h_0^-}

\def\gback{g}
\def\fpt{K_1}
\def\spt{K_2}

\def\energy{E}
\def\pressure{P}
\def\Eb{\energy}

\def\Pppz{\pressure^{(2)}_0}

\def \Ptppz{\tilde{\pressure}^{(2)}_0}

\def \Pc{P_c}
\def \mfinal{m_0{}_\natural}
\def \Pfinal{\tilde{P}_0{}_\natural}

\begin{document}
\title{Slowly rotating homogeneous masses revisited}
\author{
Borja Reina \footnote{ email: borja.reina@ehu.es}\\
Dept. of Theoretical Physics and History of Science,\\ University of the Basque Country UPV/EHU,\\
644 PK, Bilbao 48080, Basque Country, Spain}
\date{}
\maketitle
\maketitle
\begin{abstract}
Hartle's model for slowly rotating stars has been extensively used to compute equilibrium configurations of slowly rotating stars to second order in perturbation theory in General Relativity, given a barotropic equation of state (EOS). A recent study based on the modern theory of perturbed matchings show that the model must be amended to accommodate EOS's in which the energy density does not vanish at the surface of the non rotating star. In particular, the expression for the change in mass given in the original model, i.e. a contribution to the mass that arises when the perturbations are chosen so that the pressure of the rotating and non rotating configurations agree, must be modified with an additional term. In this paper, the amended change in mass is calculated for the case of  constant density stars. \end{abstract}
\section{Introduction}

Hartle's model \cite{Hartle1967} describes the axially symmetric equilibrium configuration
of a rotating isolated compact body and its vacuum exterior in perturbation theory in GR.
The interior of the body is a perfect fluid satisfying a barotropic equation of state, it does not have convective motions and rotates rigidly. The exterior is an asymptotically flat, stationary and axisymmetric vacuum and both the interior and the exterior share equatorial symmetry. The interior and exterior are matched across a timelike hypersurface that represents the boundary of the star. 

Hartle's model is built taking slow rotation perturbations around a static and spherically symmetric background. The perturbed field equations for the interior region are presented with conditions at the origin so that, once a barotropic EOS is specified, regular solutions for the fluid are found. On the other hand, the vacuum equations  can be analytically solved and the solutions compatible with asymptotic flatness are given. The solutions for both fluid and vacuum are completely determined assuming that the metric perturbations are continuous in the coordinates used at the boundary of the star, where the fluid and vacuum spacetimes are matched.

However, the  theory that provides the matching conditions in a perturbed sense  to first and second order, formulated independently of the coordinates used to describe the background configurations and also of the gauges (both spacetime and hypersurface) used to describe the perturbed configurations, is very recent \cite{Mars2005}.  
Hartle's model has been revisited in a fully consistent manner from the perspective of the modern theory of perturbed matchings in \cite{ReinaVera2014}. In particular, in that work it has been found that a function in the second order perturbations, $m_0(r)$, 
in the coordinates and gauges used in the original work \cite{Hartle1967} is not continuous, in contradiction with the implicit assumption there. It turns out that the discontinuity of the function $m_0(r)$  affects a particular outcome of the model, the change in mass $\delta M$, which is the contribution to the mass of the rotating configuration that arises when the central densities (pressures in this paper) of the non-rotating and rotating configurations are forced to agree. The jump on a radial $m_0$ function already appears implicitly in the revision of Hartle's model in \cite{Bradley_etal2007}, where the matching hypersurface is prescribed as the set of points
where the perturbed pressure vanishes in a certain gauge.

The correction to $\delta M$ found in \cite{ReinaVera2014} is proportional to the energy density evaluated at the (non rotating) surface of the star. Hence, it is negligible when the usual equations of state for neutron stars are considered, since the pressure and the energy density typically decrease together and vanish simultaneously at the surface of the star. This is the behaviour shown  by, e.g. polytropic EOS's. Nevertheless, the correction may be relevant in other EOS's for which the energy density takes a finite value at the boundary, which is precisely the case of linear equations of state, for instance those used to describe strange quark stars \cite{ColpiMiller}, or  constant density (homogeneous) stars \cite{Chandra-Miller1974}.  

In this work the particular case of homogeneous stars is studied. 
This equation of state may not be realistic in physical terms, but its study is interesting  for several reasons. First of all the spherically symmetric and static configuration can be solved analytically and hence, the perturbed field equations become simpler. 

Secondly, it is interesting to know how noticeable the correction for the change in mass might be in numerical terms. There are factors in the change in mass that are easy to estimate and, in fact, can be taken as inputs for the model, as the mass and the size of the static and spherically symmetric star. In contrast, the perturbation to the pressure is not easily estimated and the model must be solved to second order. Another important factor is the value of the energy density at the surface of the star in the nonrotating configuration and this is precisely the reason for having chosen this particular EOS. It is probably one of the most favourable cases for the correction. Thus, it is reasonable to think that the constant energy EOS may constitute a numerical bound for the amended change in mass, since for any other of the usual EOS the value of the energy density at the surface will not be as important as in the present case.  

Homogeneous stars drew the attention of Chandrasekhar and Miller  \cite{Chandra-Miller1974}, back in 1974. In that work, they used Hartle's formalism to solve the homogeneous star in the slow rotating approximation.
These have also been studied under the CMMR formalism \cite{CMMR}, a treatment based on the post-Minkowskian and small deformation approximations in \cite{CuchiLinear}, as a particular case of the linear EOS.

In this paper Chandrasekhar and Miller's work is retaken, following Hartle's setting as described in \cite{ReinaVera2014} and taking into account the correct value of $\delta M$. The results are presented in an analogous way to \cite{Chandra-Miller1974} to ease the comparison of tables and figures.

\section{Hartle's model in brief}
Hartle's model is based upon the following one parameter family of metrics \footnote{The function $m(r,\theta)$ in (\ref{gefamily}) is the same function $m(r,\theta)$ that is used in \cite{Hartle1967}. In contrast, the $m(r,\theta)$ in \cite{ReinaVera2014} corresponds to the $r e^{-\lambda(r)}m(r,\theta)$ here.}
\begin{eqnarray}
\gfamp_{\pertp} &=& -e^{\nu(r)}\left(1+2 \pertp^2 h(r,\theta)\right)dt^2 + e^{\lambda(r)}\left(1+ 2 \pertp^2 \frac{e^{\lambda(r)}}{r} m(r,\theta) \right)dr^2 \nonumber\\
&& + r^2(1+2 \pertp^2 k(r,\theta))\left[d\theta^2 + \sin ^2 \theta (d\varphi-\pertp \omega(r,\theta) dt)^2\right]\,
+ \mathcal{O}(\pertp^3), \label{gefamily} \nonumber
\end{eqnarray}
defined on the static and spherically symmetric background spacetime $(\mathcal{V}_0, g:=g_{\varepsilon=0})$ so that
\begin{equation}
\gback= -e^{\nu(r)} d{t}^ 2 + e^{\lambda(r)} d{r}^ 2+{r}^ 2(d{\theta}^2+\sin^2 \theta d {\varphi}^ 2).\label{eq:g0}
\end{equation}

The first and second order metric perturbation tensors, $\fpt= \left.\partial_\pertp g_\pertp\right|_{\pertp=0}$ and $\spt =\left. \partial_\pertp^2 g_\pertp\right|_{\pertp=0}$ take the form
\begin{eqnarray}
\fpt  &=& 
 -2r^2\, \omega(r) \sin ^2 \theta dt d\varphi \label{fopert_tensor},\\
\spt &=& \left(-4 e^{\nu(r)} h(r, \theta) + 2r^2 \sin ^2 \theta {\omega}^2(r, \theta)\right)dt^2 + \frac{4 e^{2\lambda(r)}}{r} m(r, \theta) dr^2 \nonumber\\
&&+4 r^2 k(r, \theta)
(d\theta^2 + \sin ^2 \theta d\varphi^2),\label{sopert_tensor}
\end{eqnarray}
where the functions in $\spt$ admit the decompositions $h(r,\theta) = h_0(r) + h_2(r) P_2(\cos \theta)$, $m(r,\theta) = m_0(r) + m_2(r) P_2(\cos \theta)$ and  $k(r,\theta) = k_2(r) P_2(\cos \theta)$. Note that the choice $k_0(r)=0$ fixes the second order spacetime gauge at both regions (fluid and vacuum, see \cite{ReinaVera2014}). The second order perturbation tensor as written in (\ref{sopert_tensor}) with $k_0(r)=0$ is said to be in the $k$-gauge in \cite{ReinaVera2014}.
The perturbed Einstein's field equations are imposed as follows. Consider the $\varepsilon$ family of equations $G (g_\varepsilon)_{\alpha \beta} = 8 \pi{T_\varepsilon}_{\alpha \beta}$ and take the first order field equations as the first $\varepsilon$-derivative at $\varepsilon = 0$, i.e. $\partial_\varepsilon G(g_\varepsilon)_{\alpha \beta}|_{\varepsilon=0} = \partial_\varepsilon {T_\varepsilon}_{\alpha \beta}|_{\varepsilon=0}$ and the same for successive orders.  For the interior region consider a family of perfect fluids so that $T_\varepsilon = (E_\varepsilon + P_\varepsilon)u_\varepsilon \otimes u_\varepsilon + P_\varepsilon g_\epsilon$. The unit fluid flow under the assumption of circularity and rigid rotation explicitly reads  
\begin{eqnarray}
&&\vec{u}_\varepsilon = e^{-\nu/2}\partial_t + \varepsilon \Omega \partial_\varphi + \frac{\varepsilon^2}{2} e^{-3\nu/2}(\Omega^2 g_{\varphi \varphi} + 2\Omega K_{1t\varphi} + K_{2tt}/2)\partial_t+ \mathcal{O}(\varepsilon ^3) ,
\end{eqnarray}
for some constant $\Omega$.
The energy density and pressure admit the expansions
\begin{eqnarray}
 E_\varepsilon (r,\theta) = E (r) + \frac{\varepsilon^2}{2} E^{(2)}(r,\theta),	\quad P_\varepsilon (r,\theta) = P(r) + \frac{\varepsilon^2}{2}P^{(2)}(r,\theta), \label{expansions}
\end{eqnarray}
where the first order terms in $\varepsilon$ vanish
due to the structure of the perturbed field equations to first order 
\cite{Hartle1967} (see also \cite{Bradley_etal2007,ReinaVera2014}). The field equations lead to  $E^{(2)}(r,\theta) = E_0^{(2)}(r) + E_2^{(2)}(r)P_2(\cos \theta)$ and $P^{(2)}(r,\theta) = P_0^{(2)}(r) + P_2^{(2)}(r)P_2(\cos \theta)$ \cite{Hartle1967} (see also \cite{Bradley_etal2007,ReinaVera2014}).

Consider two copies of the spacetime introduced so far, $(\mathcal{V}^+_0, g^+)$ with the corresponding family $g_\varepsilon^+$ to describe the perfect fluid region, and $(\mathcal{V}^-_0, g^-)$ with $g_\varepsilon^-$ for the vacuum region. The coordinates are labelled with subscripts in order to avoid confusion.

The background configuration for the interior region  satisfies the equations of general relativistic hydrostatics addressed in \cite{Hartle1967}
\begin{eqnarray}
\quad \frac{ d\Mf(\rf)}{d\rf} &=& 4 \pi \rf{}^2 E (\rf), \label{eq:MTOV}\\
\frac{dP(\rf)}{d\rf} &=& - \frac{(E(\rf)+P(\rf))(\Mf(\rf)+4\pi \rf{}^3 P)}{\rf(\rf -2\Mf(\rf))},\label{eq:PTOV}
\end{eqnarray}
where the functions $\Mf(\rf)$ and $P(\rf)$ are related to the metric functions $\lf (\rf)$  and $\nuf (\rf)$ by
\begin{eqnarray}
e^{-\lf (\rf)} &=& 1-\frac{2 \Mf(\rf)}{\rf}\;, \label{eq:lambdaM}\\
 \frac{d \nuf (\rf)}{d \rf} &=& -\frac{2}{E(\rf)+P(\rf)} \frac{dP(\rf)}{d\rf}. \label{eq:nuP}
\end{eqnarray}
This set of equations determines the interior configuration once the values of the energy density $E$ and pressure $P$ at the origin are given. The function $\nuf$ is determined up to an additive constant.

For the exterior vacuum $E = P = 0$, thus $\Mv(\rv)=M$ is a constant and the metric functions are given by 
\begin{eqnarray}
e^{\nuv(\rv)} = e^{-\lv (\rv)} = 1-\frac{2M}{\rv}.\label{sol:nuext}
\end{eqnarray}

These two background spacetimes $(\mathcal{V}_0^+,g^+)$ and $(\mathcal{V}_0^-,g^-)$ are matched across  diffeomorphic boundaries, $\Sb :=\Sigma_0^\pm$. The most general matching preserving the symmetries \cite{Vera2002} (spherical and static) can be cast in parametric form in terms of three coordinates $\{\tau, \phi, \vartheta\}$ on $\Sigma_0$ as  
\begin{equation}
\Sf = \{\tf = \tau, \phif = \phi, \rf = \Rf, \thetaf = \vartheta\}\;, \Sv = \{\tv = \tau, \phiv = \phi, \rv = \Rv, \thetav = \vartheta\},
\end{equation}
for some constants $\Rf$, $\Rv$ without loss of generality, and must satisfy (e.g. \cite{ReinaVera2014})
\begin{equation}
R\equiv \Rf=\Rv\;,\;\; [\lambda] = 0\;,\;\; [\nu]=0\;,\;\; [\nu']=0, \label{sphst_matching}
\end{equation}
where $[f]$ is the difference of a function $f$ at the fluid region and the value of $f$ at the vacuum region evaluated on $\Sb$, i.e. $[f] := f^{+}|_{\Sb} - f^{-}|_{\Sb}$.
The first condition establishes that the constants $R^{\pm}$ must agree, the condition $[\lambda] = 0$ relates the constant $M$ of the vacuum solution with the mass of the fluid $\Mf(R)=M$, $[\nu]=0$ fixes the value of the function at the origin and $[\nu']=0$ then implies that the pressure must vanish at the surface of the star $P(R)=0$. The constant $R$ is called $a$ in \cite{Hartle1967, ReinaVera2014}.

With the background configuration already matched, the first order perturbations can be studied. In order to cast the first order field equations in a compact form, an auxiliary  function $j$ is defined as
\begin{equation}
j := e^{-(\nu + \lambda)/2}\,,\qquad \jv=1.
\end{equation}

In the first order perturbation tensor (\ref{fopert_tensor}), only the function $\omega (r)$ is involved. In fact, it is convenient to work with the function $\tilde{\omega}(r) := \Omega - \omega (r)$ that satisfies the single field equation  \cite{Hartle1967}
\begin{equation}
\frac{1}{r^3}\frac{d}{dr}\left( r^4 j \frac{d\tilde{\omega}}{dr}\right) + 4 \frac{dj}{dr} \tilde{\omega} = 0. \label{eqfo}
\end{equation}
This equation is integrated from the origin outwards, with conditions that ensure that the solution is regular at the origin. Thus, there is only one parameter that must be provided to completely determine the interior solution. In the present work it is chosen to be the value of $\of$ at the origin, denoted by $\of_c$, although it is common in the literature to specify others, like the critical angular velocity for which rotational shedding occurs \cite{HartleThorne1968}.

Equation (\ref{eqfo}) in the vacuum region holds for $j=1$, which leads to 
\begin{equation}
\ov (\rv) = \Omega - \frac{2J}{{\rv}^3}, \label{sol:omega_vac}
\end{equation}
given asymptotic flatness.
Choosing the spacetime gauge to first order in the fluid region conveniently, the functions $\of$ and $\ov$ satisfy the matching conditions $[\tilde{\omega}]=[\tilde{\omega}']=0$ \cite{ReinaVera2014}. Hence the constant $\Omega$ is indeed the rotation of the fluid measured by a distant observer (see \cite{ReinaVera2014} for a deeper discussion) . This constant along with the angular momentum of the star $J$ are thus determined by
\begin{eqnarray}
J = \frac{1}{6} R^4 \left( \frac{d\of}{d\rf}\right)_{\rf=R},\qquad  \Omega =\of(R) + \frac{2J}{R^3}. \label{JOmega}
\end{eqnarray}
The moment of inertia $I$ is defined in terms of these two constants as $I := J/\Omega$.

Regarding the second order perturbations, the change in  mass involves the $l=0$ sector alone and this can be studied independently from the $l=2$ sector \cite{Hartle1967}. The functions involved in the $l=0$ sector are thus $m_0$ and $h_0$. Although in \cite{Bradley_etal2007,Hartle1967} and \cite{ReinaVera2014} the approaches taken to describe the deformation of the star are different at a fundamental level,  the deformation is described in the three references  by the value that the perturbation of the pressure $\Pppz$ takes at the surface of the non rotating star. 
For convenience, this function is scaled as
\begin{equation}
\Ptppz(\rf) := \frac{\Pppz(\rf)}{2(E(\rf)+P(\rf))}.
\end{equation} 
In the interior region the function $\hf$ is replaced by $\Pppz$ and the field equations provide a system of  first order inhomogeneous ODE's for the set $\{\mf, \Ptppz\}$. It reads \cite{Hartle1967,ReinaVera2014} 
\begin{eqnarray}
 \frac{d \mf}{d\rf} &=& 4 \pi \rf^2 \frac{dE}{dP} (E+P)\Ptppz +\frac{1}{12}{\jf}^2 {\rf}^4 \left( \frac{d\of}{d\rf} \right)^2-\frac{2}{3}{\rf}^3
\jf \frac{d\jf}{d\rf}{\of}{}^2,  \label{eq:m0r}\\
\frac{d \Ptppz}{d\rf} &=& -\frac{4\pi (\Eb + P)\rf{}^2}{\rf-2 \Mf(\rf)}\Ptppz 
- \frac{\mf \rf{}^2 }{(\rf-2\Mf(\rf))^2}\left( 8\pi P + \frac{1}{\rf {}^2}\right) \nonumber \\
&& + \frac{\rf{}^4 \jf{}^2}{12(\rf-2 \Mf(\rf))} \left(\frac{d \of}{d\rf}\right)^2 + \frac{1}{3} \frac{d}{d\rf}\left(\frac{\rf{}^3\jf{}^2 \of{}^2}{\rf-2 \Mf(\rf)} \right).\label{eq:P20} 
\end{eqnarray}
The equations must be integrated with conditions that ensure  regularity at the origin.
In addition, it is imposed that the central pressure of the non rotating configuration  is preserved in the rotating model, so that $\Ptppz(0) = 0$ (see \cite{Bradley_etal2007} for a deeper discussion).
The system admits the \emph{hydrostatic equilibrium} first integral \cite{Hartle1967}, which allows for the determination of $\hf$. Explicitly,  
\begin{equation}
\Ptppz + \hf - \frac{1}{3}\rf{}^2 e^{-\nuf}\of{}^2 = \gamma, \nonumber
\end{equation}
where $\gamma$ is a constant. Note that, once $\Ptppz (0)=0$, $\gamma$ equals the value of $\hf$ at the origin. 

The set of functions that determine the exterior configuration to second order is $\{\mv, \hv\}$. The field equations are  (\ref{eq:m0r}),  with  $\jv=1$ and $\ov$ given by (\ref{sol:omega_vac}), and  the following first order equation for $\hv$ 
\begin{equation}
\frac{d\hv}{d\rv} = \frac{\mv}{(\rv - 2M)^2} - \frac{3J^2}{\rv^4(\rv-2M)} . \nonumber
\end{equation}
The asymptotically flat vacuum solution thus reads \cite{Hartle1967}
\begin{eqnarray}
\mv(\rv) &=& \delta M -\frac{J^2}{\rv{}^3},\label{sol:m0ext}\\
\hv(\rv) &=& -\frac{\delta M }{\rv-2M} + \frac{J^2}{\rv{}^3(\rv-2M)},\nonumber
\end{eqnarray}
for some arbitrary constant $\delta M$. This constant is identified as the change in mass. 

At this point the interior solution $\{\mf, \Ptppz\}$ is completely determined,
while the exterior solution $\mv$ (\ref{sol:m0ext}) is determined up to  $\delta M$. In order to fix it the interior and the exterior solutions must be related at $\Sb$. The full second order matching is presented in \cite{ReinaVera2014} but in here  only the result regarding the functions $m_0^\pm$ is needed, which on $\Sigma_0$ satisfy
\begin{equation}
[m_0] = -4 \pi \frac{R^3}{M}(R-2M) E(R) \Ptppz(R). \label{m0_matching}
\end{equation}
This matching condition thus fixes the constant $\delta M$  as
 \begin{eqnarray}
  \label{eq:deltaM}
  && \delta M = \mf(R)+\frac{J^2}{R^3} + 4\pi\frac{R^3}{M}(R-2M)E(R) \tilde{P}_0^{(2)}(R).
 \end{eqnarray}
In order to compare the results here with \cite{Chandra-Miller1974} the  change in mass is split into two contributions: the one given originally in  reference \cite{Hartle1967}  and used in \cite{Chandra-Miller1974}, $\delta M^{(O)}:=\mf(R)+\frac{J^2}{R^3}$ and the correction found in \cite{ReinaVera2014}, $\delta M^{(C)}=4\pi\frac{R^3}{M}(R-2M)E(R) \tilde{P}_0^{(2)}(R)$.

\section{Homogeneous stars}

When the energy density $E$ is constant, the equations of structure (\ref{eq:lambdaM}), (\ref{eq:nuP})  that govern the (static, spherically symmetric) background configuration admit an analytical solution. In terms of the  constant density $E$ and the central pressure $\Pc$, that solution is given by 
\begin{eqnarray}
\frac{P + \frac{E}{3}}{P+E} &=&\frac{\Pc + \frac{E}{3}}{\Pc+E}\sqrt{1-\frac{8\pi E {\rf}^2}{3}},\label{sol:pressure}\\
e^{-\lf(\rf)} &=& 1- \frac{2 \Mf(\rf)}{\rf}\,, \qquad \Mf(\rf) = \frac{4\pi}{3} E \rf{}^3 \; ,\label{sol:lambda}\\
e^{\nuf(\rf)/2} &=& e^{\nuf(0)/2}\left ( -1+\frac{\Pc + \frac{E}{3}}{\Pc+E}\sqrt{1-\frac{8\pi E \rf{}^2}{3}}\right)\left(-1 + \frac{\Pc + \frac{E}{3}}{\Pc+E}\right )^{-1} \label{sol:nu}.
\end{eqnarray}
This solution stands for the whole interior region, i.e. from $\rf=0$ to $\rf = R$. 
The vacuum solution is given by (\ref{sol:nuext}) and extends from $\rf = R$ to infinity. The two solutions are related by means of the matching conditions for the background configuration (\ref{sphst_matching}). The ``continuity'' of $\lambda$ and $\nu'$ implies that $M = 4 \pi E R^3/3$ and $P(R)=0$.

With the background configuration already matched, it is convenient to change from the \emph{interior} parameters  $\{E, \Pc\}$ to the \emph{exterior} parameters $\{M ,R\}$. Thence the solution (\ref{sol:pressure}) takes the form as given in  \cite{Gravitation, Wald}. In order to present the results as in \cite{Chandra-Miller1974}, the exterior parameters still have to be scaled with the Schwarzschild radius $R_S:=2M$ so that they become $\{R_S, R/R_S\}$. Inverting the relation between the parameters one finds
\begin{eqnarray}
E = \frac{3}{8\pi R_S^2}\left(\frac{R}{R_S} \right)^{-3}\,,\qquad \Pc = \frac{3}{8\pi R_S^2} \left(\frac{R}{R_S} \right)^{-3} \frac{1-\sqrt{1- \left(\frac{R}{R_S} \right)^{-1}}}{3\sqrt{1- \left(\frac{R}{R_S} \right)^{-1}}-1}.
\label{sol:pressure2}
\end{eqnarray}
Equation (\ref{sol:pressure2}) implies a constraint on the background exterior parameters. In order to keep the central pressure finite, the following inequality must hold \cite{Chandra-Miller1974}
\begin{equation}
0\leq \frac{9}{4} M < R \Leftrightarrow 0\leq R < \sqrt{\frac{1}{3\pi E}}. \label{masslimit}
\end{equation} 

In order to write the field equations for the fluid  as in \cite{Chandra-Miller1974},  two constants that replace   $E$ and  $R$ must be introduced as
\begin{equation}
\alpha := \sqrt{\frac{3}{8 \pi E}}, \qquad \kappa :=3 \sqrt{1-\frac{R^2}{\alpha^2}}-1 . \label{def:alphakappa}
\end{equation}
The constant  $\alpha$ is  related to the Schwarzschild radius $R_S$ by
\begin{equation}
\alpha = R_S \left(\frac{R}{R_S}\right)^{3/2}. \label{alpha_to_RS}
\end{equation}
Hence, given any function in units of $\alpha$, it can be easily converted to units of the Schwarzschild radius by simply scaling it with the proper factor  $R/R_S$. 
Following the conventions in \cite{Chandra-Miller1974}, the radial coordinate in the interior region is finally substituted by 
\begin{equation}
x := 1 - \sqrt{1-\left(\frac{\rf}{\alpha}\right)^2}.
\end{equation}
This change is well defined once  the inequality (\ref{masslimit}) holds. The domain of definition of this coordinate is  $x \in [0,2/3)$. The radius of the star is denoted by $X$, which in terms of $\kappa$ is expressed as $X = (2-\kappa)/3$.  Finally, in terms of $x$ the auxiliary function $\jf$ reads $\jf = 2(1-x)/(\kappa + x)$.

Considering the background solution (\ref{sol:pressure})-(\ref{sol:nu}) and the definitions introduced so far, the first order field equation (\ref{eqfo}) for the interior region is written in terms of the radial coordinate $x$ as
\begin{equation}
-x(2-x)(x+\kappa) \frac{d^2 \of}{dx^2}+(4x^2-x(3-5\kappa)-5\kappa)\frac{d \of}{dx} + 4(1+\kappa)\of(x)=0.\label{eq:omegax} 
\end{equation}
The behaviour of $\of$ near the origin is  fixed by
\begin{eqnarray}
\of = \of_c \left( 1+ \frac{4(1+\kappa)}{5 \kappa }x\right) + O(x^2). \label{condition:omega}
\end{eqnarray}

The second order field equations (\ref{eq:m0r}) and (\ref{eq:P20}) are  written in terms of the radial coordinate $x$ as \cite{Chandra-Miller1974}
\begin{eqnarray}
\alpha^{-3} \frac{d \mf}{dx} &=& \frac{(1-x) ((2-x) x)^{3/2}}{(\kappa +x)^2}  \left(\frac{1}{3}(2-x) x \left(\frac{d \of}{dx}\right)^2 + \frac{8 (\kappa +1)}{3(\kappa + x)} {\of}{}^2\right),\label{eq:m0x}\\
\alpha^{-2}\frac{d \Ptppz}{dx} &=& -\frac{\kappa +1}{(1-x) (\kappa +x)} \alpha^{-2} \Ptppz - \frac{2+(\kappa + 1)(1-x)-3(1-x)^2}{(\kappa + x)(1-x)^2 x^{3/2}(2-x)^{3/2} } \alpha^{-3}\mf \label{eq:p2x}\\
&&+\frac{8 (2-x) x}{3 (\kappa +x)^2} \of \frac{d\of}{dx} -\frac{8 (\kappa  (x-1)+x)}{3  (\kappa +x)^3} \of{}^2 +\frac{ (2-x)^2 x^2}{3  (1-x) (\kappa +x)^2} \left(\frac{d \of}{dx} \right)^2. \nonumber
\end{eqnarray}
Regularity at the origin demands  \cite{Chandra-Miller1974}
\begin{eqnarray}
\frac{\mf}{\alpha ^{3} \of_c{}^2}(x\rightarrow 0 )&=& \frac{32 \sqrt{2}  (\kappa +1) x^{5/2}}{15  \kappa ^3}   + O(x^{7/2}),\label{condition:m0}\\
\frac{\Ptppz}{\alpha ^2\of_c{}^2 } (x\rightarrow 0 )&=& \frac{8  x}{3  \kappa ^2} + O(x^2).\label{condition:p20}
\end{eqnarray}

Once the values of the functions $\mf$ and $\Ptppz$ in $x=X$ are found, the change in mass is calculated using expression (\ref{eq:deltaM}).
In order to present the numerical results, it is convenient to express $\delta M$ divided by the mass of the background configuration $M$, and in units of $J^2/R_S^4$. It reads

\begin{equation}
\frac{\delta M}{M} = \frac{J^2}{R_S^4} \left\lbrace \left. 2 \frac{\mf}{(J^2/R_S^3)}\right|_{\Sigma_0}  + 2 \left(\frac{R}{R_S}\right)^{-3} +\left. 6 \left(\frac{R}{R_S}-1\right)\frac{\Ptppz}{(J^2 /R_S^4)}\right|_{\Sigma_0}\right \rbrace. \label{excessmass2}
\end{equation}

Note that the field equations for the first (\ref{eq:omegax}) and second  order (\ref{eq:m0x}), (\ref{eq:p2x}) and the conditions at the origin (\ref{condition:omega}) and (\ref{condition:m0}),  (\ref{condition:p20}) are formulated for $\of/\of_c$, $\mf/\alpha^3 \of_c{}^2$ and $\Ptppz/\alpha^2 \of_c{}^2$, and thus, depend only on one free parameter, $\kappa$, from the background configuration. This parameter $\kappa$ is in fact determined by the ratio of the radius of the spherical star to the Schwarzschild radius $R/R_S$ by means of (\ref{def:alphakappa}) and (\ref{alpha_to_RS}). Hence the model is solved just specifying a value of $R/R_S$. A sequence of models with different values of $R/R_S$ is explored and the results for the first and second order are presented below.

The first order results are shown in Table \ref{table:firstorder}, which includes the moment of inertia and the value $\of/(J/R_S^3)|_{\Sigma_0}$ for different values of $R/R_S$. For a solid sphere in the Newtonian regime, the normalized momentum of inertia $i:=I/MR^2$ takes the value $2/5$, which is achieved asymptotically (see  Table \ref{table:firstorder}). The comparison with a sphere makes sense because the first order perturbations do not change the shape of the star. Hence, the deviation between this value and the values shown in Table \ref{table:firstorder} is an effect of the twisted geometry. These results  \footnote{Values for $R/R_S = 9/8$ are not shown because although the perturbations can still be solved \cite{Chandra-Miller1974}, the background solution is not regular.} fully agree with those presented in \cite{Chandra-Miller1974}.

\begin{table}[h]
\centering
\begin{tabular}{ |l | c | c| c|}
  \hline                       
  $R/R_S$ & $I$ & $i$ & $\tilde{\omega}(R_0)$ \\
  \hline
  $1.15$ & 0.5105 & 0.7720 & 0.64391\\
  $1.2$ & 0.5248 & 0.7289 & 0.74818\\
  $1.3$ & 0.5657 & 0.6695 & 0.85741\\
  $1.4$ & 0.6171 & 0.6296 & 0.89174\\
  $1.5$ & 0.6758 & 0.6007 & 0.88714\\
  $1.6$ & 0.7406 & 0.5786 & 0.86205\\
  $1.7$ & 0.8106 & 0.5610 & 0.82652\\
  $1.8$ & 0.8856 & 0.5467 & 0.78622\\
  $1.9$ & 0.9653 & 0.5348 & 0.74441\\
  $2.0$ & 1.049 & 0.5247 & 0.70294\\
  $2.5$ & 1.534 & 0.4910 & 0.52376\\
  $3.0$ & 2.123 & 0.4717 & 0.39700\\
  $4.0$ & 3.604 & 0.4505 & 0.24623\\
  $5.0$ & 5.487 & 0.4390 & 0.16625\\
  $10.0$ & 20.91 & 0.4182 & 0.045821\\
  $20.0$ & 81.77 & 0.4088 & 0.011980\\
  $35.0$ & 248.1 & 0.4050 & 0.0039848\\
  $50.0$ & 504.3 & 0.4035 & 0.0019668\\
  $100.0$ & 2009 & 0.4017 & 0.00049585\\
  \hline
\end{tabular}
\caption{$I$ in units of $R_S^3$, $i:=I/MR^2$ and $\tilde{\omega}(R_0)/(J/R_S^3)$ for some values of $R/R_S$. \label{table:firstorder}}
\end{table}

\begin{figure}[top]
    \centering
\includegraphics[scale=0.70]{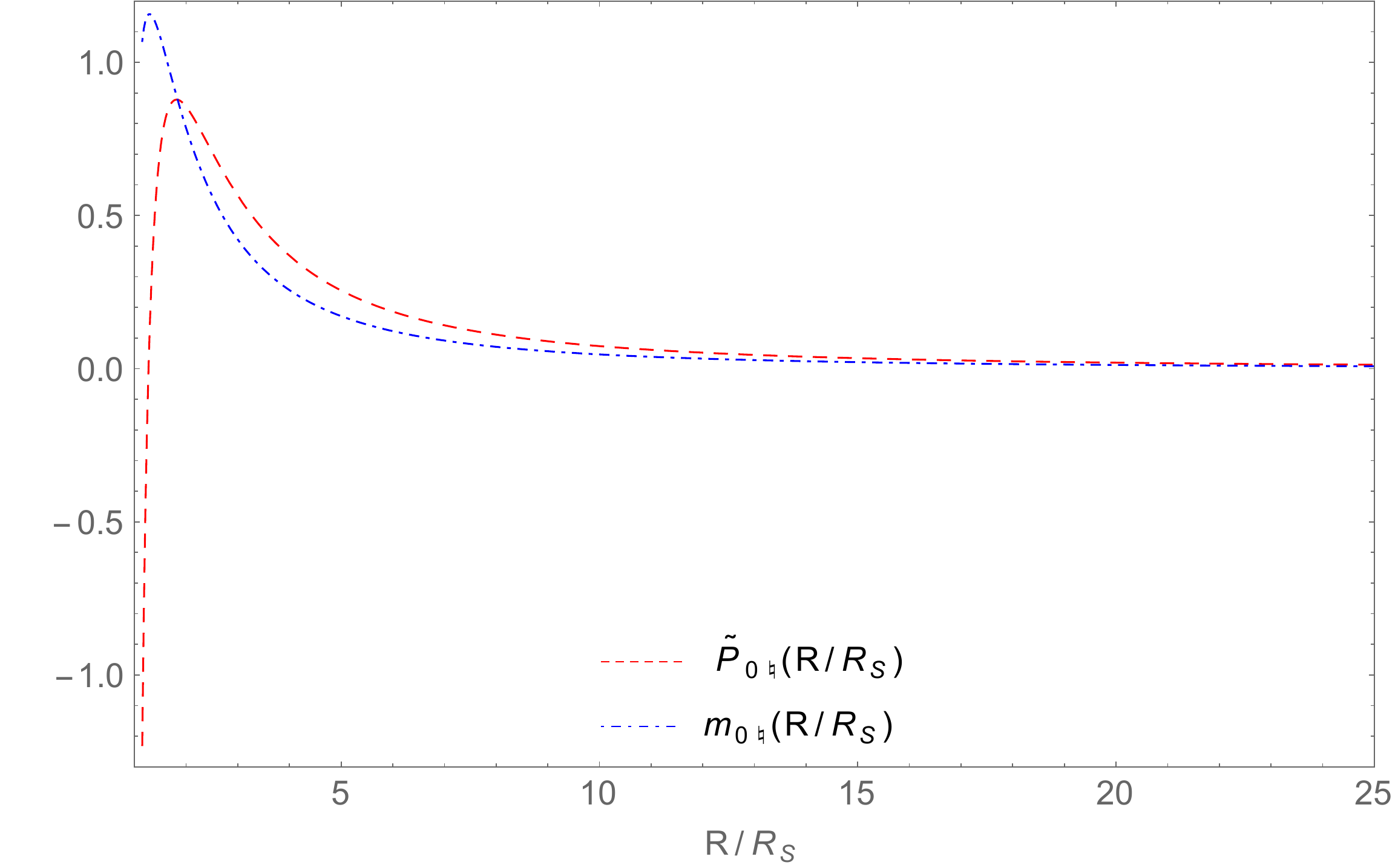}
\caption{The perturbation to the pressure to second order $\Pfinal(R/R_S)$  and $\mfinal(R/R_S)$.
 \label{m0p20}}
\end{figure}

\begin{figure}[bottom]
    \centering
\includegraphics[scale=0.70]{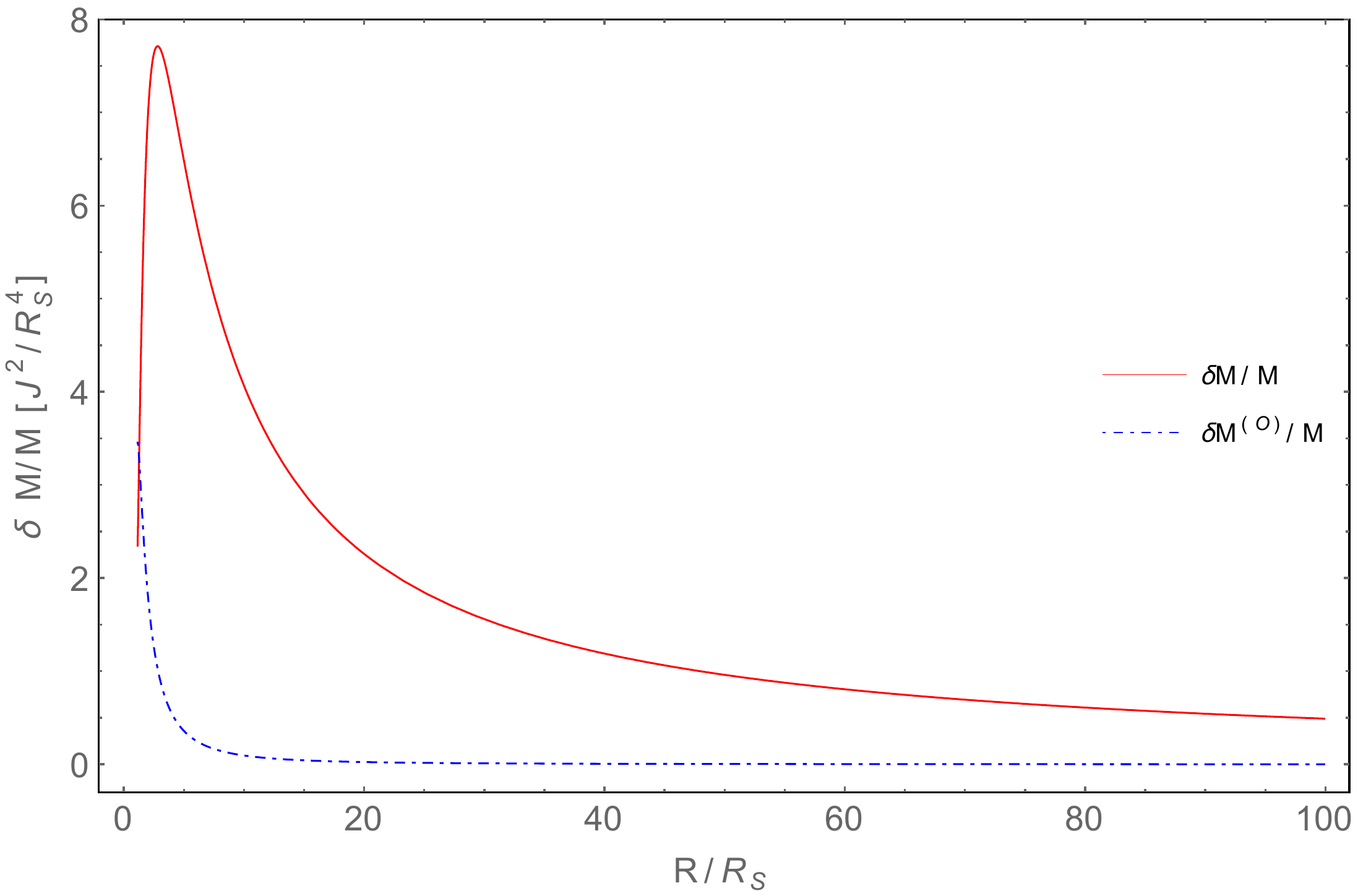}
\caption{The original  and the amended  changes in  mass versus the normalized radius of the static star $R/R_S$. \label{plotaco}}
\end{figure}

The numerical results for the second order are summarized in Table \ref{table:secondorderl0} and Figures \ref{m0p20} and \ref{plotaco}.
\begin{table}[h!]
\centering
\begin{tabular}{ |l | c|c|c | c| c|c|}
  \hline                      
  $R/R_S$ &  $\delta M^{(O)}/M$ & $\delta M/M$ & $|\delta M^{(C)}|/\delta M$\\
  \hline
  $1.15$ &  3.454 & 2.348 & 0.4711\\
  $1.2$ &  3.412 & 2.725 & 0.2524\\
  $1.3$ &  3.225 & 3.506 & 0.0803\\
  $1.4$ &  2.993 & 4.246 & 0.2952\\
  $1.5$ &  2.757 & 4.904 & 0.4379\\
  $1.6$ &  2.533 & 5.474 & 0.5372\\
  $1.7$ &  2.327 & 5.954 & 0.6091\\
  $1.8$ &  2.140 & 6.355 & 0.6632\\
  $1.9$ &  1.971 & 6.684 & 0.7051\\
  $2.0$ &  1.819 & 6.951 & 0.7383\\
  $2.5$ &  1.259 & 7.631 & 0.8350\\
  $3.0$ &  0.9163 & 7.690 & 0.8808\\
  $4.0$ &  0.5440 & 7.173 & 0.9242\\
  $5.0$ &  0.3588 & 6.482 & 0.9446\\
  $10.0$ &  0.09493 & 4.065 & 0.9766\\
  $20.0$ &  0.02437 & 2.259 & 0.9892\\
  $35.0$ &  0.008049 & 1.349 & 0.9940\\
  $50.0$ &  0.003966 & 0.9608 & 0.9959\\
  $100.0$ &  0.001008 & 0.4901 & 0.9979\\
  \hline
\end{tabular}
\caption{The change in mass $\delta M^{(O)}/M$ typical from the literature, the amended change in mass $\delta M/M$ and the fraction of the correction with respect to the total change in mass are presented for different values of $R/R_S$.\label{table:secondorderl0}}
\end{table} 
 In Figure \ref{m0p20} $\mf$ in units of $J^2/R_S^3$ and $\Ptppz$ in units of $J^2/R_S^4$ at $\rf = R$ are shown as functions of $R/R_S$, i.e. $\mf/(J^2/R_S^3)|_{\Sigma_0}(R/R_S)$ and $\Ptppz/(J^2/R_S^4)|_{\Sigma_0}(R/R_S)$ respectively. In order not to overwhelm the notation in the subsequent discussion, let us  refer to these two previous functions as $\mfinal (R/R_S)$ and $\Pfinal(R/R_S)$ respectively. As mentioned in \cite{Chandra-Miller1974} these are not monotonic functions and they both present a maximum, the first one at $R/R_S \sim 1.29$ and the second  at $R/R_S \sim  1.82$. Note that the function $\Pfinal$ is negative for small values of $R/R_S$ and this implies that the average deformation of the star to second order \footnote{The relation of the pressure and the shape of the star is addressed in \cite{Hartle1967}.} can be either negative or positive, so that the star may show either contraction or expansion to second order depending on the background parameters. It is worth noting that $\mfinal$ and $\Pfinal$ attain values of the same order and the ratio is about 1.63 for big values of $R/R_S$. For a more detailed discussion we refer the reader to the original work \cite{Chandra-Miller1974}.

Finally, the results including the corrected change in mass and their comparison with those presented in \cite{Chandra-Miller1974} are shown in Table \ref{table:secondorderl0} and Figure \ref{plotaco}. In Table \ref{table:secondorderl0} some values of the change in mass as a function of $R/R_S$ are presented. In the second column the value of $\delta M^{(O)}/M$, which corresponds to the $\delta M/M$ given in \cite{Chandra-Miller1974}, is shown, whereas the third column includes the correct change in mass (\ref{eq:deltaM}) and, lastly, the fourth column shows the fraction of the change in mass that corresponds to the correction. In fact, the  correction becomes the dominant contribution in $\delta M$ as the quotient $R/R_S$ increases.
The behaviour of $\delta M/M$ is shown for a wide range of the variable $R/R_S$ in Figure \ref{plotaco}. $\delta M/M$  presents a  maximum at $R/R_S \sim 2.81$ and then decreases more slowly than the $\delta M^{(O)}/M$ presented in \cite{Chandra-Miller1974}, which decays monotonically. The amended $\delta M/M$ and $\delta M^{(O)}/M$ agree for $R/R_S \sim 1.27158$.   

There is a combination of two facts that makes the correction $\delta M^{(C)}$ not only noticeable, but also dominant in homogeneous stars. On the one hand, as shown in Figure \ref{m0p20}, $\mfinal$ and $\Pfinal$ are quantities of the same order. On the other hand, in formula (\ref{excessmass2}), the coefficient with  $\Pfinal$ scales linearly with $R/R_S$.
These results definitely reveal that  the correction to the change in mass is important in homogeneous stars and it can not be neglected at all. Again, the correction to $\delta M$ depends crucially on the value of the energy density at the surface of the static and spherically symmetric configuration, and hence, it does not contribute, in principle, to $\delta M$ in the usual EOS's for neutron stars. However  actual models of neutron stars show a  residual value associated with the numerical error present when the radius of the star is determined by $P(R)=0$. Therefore, the correction to the change in mass  might be relevant in order to improve the precision in the determination of the mass of the rotating configuration. This fact regarding the usual EOS's for neutron stars and other cases of interest are currently under study.

\section*{Acknowledgements}
I am grateful to Raül Vera for his valuable comments and careful reading of the manuscript. I would also like to thank  Michael Bradley for drawing my attention to reference \cite{Chandra-Miller1974} and Irene Sendra for helpful suggestions. I acknowledge financial support from the Basque Government grant BFI-2011- 250.
\bibliography{references}{}
\bibliographystyle{plain}

\newpage

\end{document}